\documentclass[british,twocolumn,prb,,superscriptaddress]{revtex4-1}
\usepackage[T1]{fontenc}
\usepackage[latin9]{inputenc}
\usepackage{babel}
\usepackage{amstext}
\usepackage{amssymb}
\usepackage{graphicx}
\usepackage[unicode=true,pdfusetitle,
 bookmarks=true,bookmarksnumbered=false,bookmarksopen=false,
 breaklinks=false,pdfborder={0 0 1},backref=false,colorlinks=false]
 {hyperref}
\usepackage{breakurl}

\makeatletter
 \@ifundefined{textcolor}{}
 {%
  \definecolor{BLACK}{gray}{0}
  \definecolor{WHITE}{gray}{1}
  \definecolor{RED}{rgb}{1,0,0}
  \definecolor{GREEN}{rgb}{0,1,0}
  \definecolor{BLUE}{rgb}{0,0,1}
  \definecolor{CYAN}{cmyk}{1,0,0,0}
  \definecolor{MAGENTA}{cmyk}{0,1,0,0}
  \definecolor{YELLOW}{cmyk}{0,0,1,0}
  }

\makeatother

\begin{document}

\title{Superconducting gap in BaFe$_{2}$(As$_{1-x}$P$_{x}$)$_{2}$ from
temperature dependent transient optical reflectivity }

\author{A. Pogrebna }

\affiliation{Complex Matter Dept., Jozef Stefan Institute, Jamova 39, Ljubljana,
SI-1000, Ljubljana, Slovenia }

\affiliation{Jožef Stefan International Postgraduate School, Jamova 39, SI-1000
Ljubljana, Slovenia}

\author{T. Mertelj }

\email{tomaz.mertelj@ijs.si}

\selectlanguage{british}%

\affiliation{Complex Matter Dept., Jozef Stefan Institute, Jamova 39, Ljubljana,
SI-1000, Ljubljana, Slovenia }

\affiliation{CENN Nanocenter, Jamova 39, Ljubljana SI-1000, Slovenia}

\author{Z. R. Ye}

\affiliation{State Key Laboratory of Surface Physics, Key Laboratory of Micro
and Nano Photonic Structures (Ministry of Education), Department of
Physics, and Advanced Materials Laboratory, Fudan University, Shanghai
200433, China}

\author{D. L. Feng}

\affiliation{State Key Laboratory of Surface Physics, Key Laboratory of Micro
and Nano Photonic Structures (Ministry of Education), Department of
Physics, and Advanced Materials Laboratory, Fudan University, Shanghai
200433, China}

\author{D. Mihailovic}

\affiliation{Complex Matter Dept., Jozef Stefan Institute, Jamova 39, Ljubljana,
SI-1000, Ljubljana, Slovenia }

\affiliation{Jožef Stefan International Postgraduate School, Jamova 39, SI-1000
Ljubljana, Slovenia}

\affiliation{CENN Nanocenter, Jamova 39, Ljubljana SI-1000, Slovenia}
\begin{abstract}
Temperature and fluence dependence of the 1.55-eV optical transient
reflectivity in BaFe$_{2}$(As$_{1-x}$P$_{x}$)$_{2}$ was measured
and analysed in the low and high excitation density limit. The effective
magnitude of the superconducting gap of $\sim$5 meV obtained from
the low-fluence-data bottleneck model fit is consistent with the ARPES
results for the $\gamma$-hole Fermi surface. The superconducting-state
nonthermal optical destruction energy was determined from the fluence
dependent data. The in-plane optical destruction energy scales well
with $T_{\mathrm{c}}^{2}$ and is found to be similar in a number
of different layered superconductors.
\end{abstract}

\date{\today}

\maketitle

\section{Introduction}

In iron pnictides the superconductivity appears from parent spin density
wave (SDW) antiferromagnetic state as a result of doping or application
of external pressure/strain. In BaFe$_{2}$(As$_{1-x}$P$_{x}$)$_{2}$
the superconducting (SC) state is induced by means of the chemical
strain induced by the isovalent substitution of arsenic by phosphorous.
In optimally doped BaFe$_{2}$(As$_{0.7}$P$_{0.3}$)$_{2}$, where
the critical temperature reaches $T_{\mathrm{c}}=30$ K, the SC gaps
where thoroughly analysed by means of angle-resolved photoemission
spectroscopy\cite{zhangYe2012} (ARPES). Contrary to the optimally
doped Ba(Fe$_{1-x}$Co$_{x}$)$_{2}$As$_{2}$, where there is little
indication of nodes\cite{TerashimaSekiba2009,ReidTanatar2010}, nodes
in the SC gap were clearly observed\cite{KimHirschfeld2010,zhangYe2012}
in BaFe$_{2}$(As$_{0.7}$P$_{0.3}$)$_{2}$.

To study a possible effect of the nodes on the photoexcited quasiparticle
relaxation and to supplement the ARPES\cite{zhangYe2012} results
on the SC gap sizes with a more bulk-sensitive technique we therefore
conducted a systematic 1.55-eV optical transient reflectivity study
in optimally doped BaFe$_{2}$(As$_{0.7}$P$_{0.3}$)$_{2}$. It was
found that similarly to the SC cuprates the nodes do not suppress
the formation of the Rothwarf-Taylor\cite{RothwarfTaylor1967,KabanovDemsar2005}
relaxation bottleneck. The behaviour is consistent with previous time-resolved
optical spectroscopy data\cite{TorchinskyMcIver2011,StojchevskaMertelj2012}
in related electron doped BaFe$_{2}$As$_{2}$ (Ba-122), together
with the presence of the normal-state pseudogap. The effective SC
gap obtained from the low-fluence linear-response data is consistent
with the ARPES results.\cite{zhangYe2012}

\section{Experimental details}

Single crystals of BaFe$_{2}$(As$_{0.7}$P$_{0.3}$)$_{2}$ were
grown from self flux at Fudan University.\cite{zhangYe2012} A sample
from the same batch as the one used for our experiment showed the
onset of superconductivity at $T_{C}=30$ K as determined by the SQUID
susceptibility and electric transport measurements. For optical measurements
the crystal was glued onto a copper sample holder and cleaved by a
razor blade before mounting into an optical liquid-He flow cryostat.

Measurements of the transient photoinduced reflectivity, $\Delta R/R$,
were performed using the standard pump-probe technique, with 50 fs
optical pulses from a 250-kHz Ti:Al$_{2}$O$_{3}$ regenerative amplifier
seeded with an Ti:Al$_{2}$O$_{3}$ oscillator. We used the pump and
probe photons with the laser fundamental ($\hbar\omega_{\mathrm{P}}=1.55$
eV) photon energy. An analyser oriented perpendicularly to the pump
beam polarization was used for rejection of the pump scattered light.
The pump and probe beams were nearly perpendicular to the cleaved
sample surface (001) with polarizations perpendicular to each other.
The beam diameters were calibrated by measuring the transmission through
a set of different size pinholes mounted at the sample position.

\section{Results }

\begin{figure}
\includegraphics[angle=-90,width=1\columnwidth]{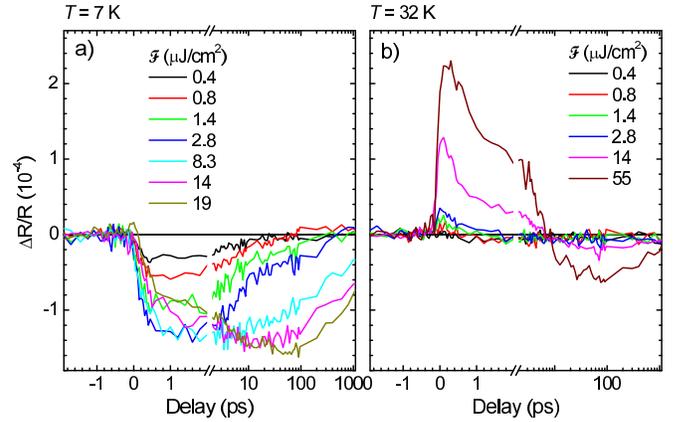}

\caption{\label{fig:DrvsF} Photoinduced reflectivity transients $\Delta R/R$
in BaFe$_{2}$(As$_{0.7}$P$_{0.3}$)$_{2}$ measured in the superconducting
state (a) and normal state (b) as a function of the pump fluence. }
\end{figure}

In Fig. \ref{fig:DrvsF} we show the pump fluence ($\mathcal{F}$)
dependence of the transient reflectivity in the SC state ($T=7$ K)
compared with the normal state transients measured just above $T_{\mathrm{c}}$
($T=32$ K). The signals show no significant pump and probe polarization
dependence. In the SC state the amplitude of the signal depends linearly
on $\mathcal{F}$ at low fluences and saturates with increasing fluence
above $\sim3$$\mu$J/cm$^{2}$. The normal-state response shown in
Fig. \ref{fig:DrvsF} (b) appears much weaker at low fluencies with
a different sign and a faster relaxation time in comparison to the
SC response. At high fluencies, above $\sim10$$\mu$J/cm$^{2}$,
the magnitude of the normal state response becomes comparable to the
SC-response magnitude due to saturation of the SC-response.

\begin{figure}
\includegraphics[angle=-90,origin=c,width=1\columnwidth]{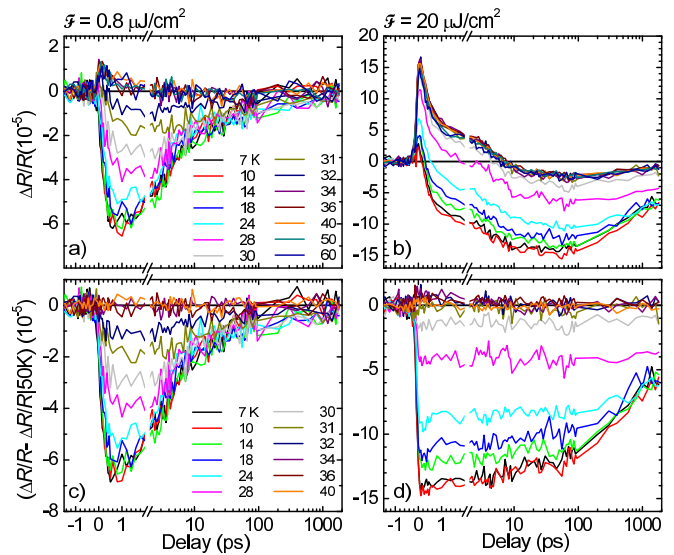}

\caption{Temperature dependence of the transient reflectivity $\Delta R/R$
at low (a) and high (b) fluence. Note that the data above the critical
temperature collapse on a single curve in a wide temperature range
at both fluencies. The superconducting state response obtained by
subtracting the average of the transients from the 34 K - 50 K interval
at low (c) and high (d) fluence.}
\label{fig:DrvsT} 
\end{figure}

In Fig. \ref{fig:DrvsT} we show the temperature dependence of the
transient reflectivity at two selected fluences. The lower was chosen
to be in the $\mathcal{F}-$linear SC-response region in most of the
$T$ range while the higher corresponds to the strongly saturated
SC-response fluence. In both cases the data in the normal state collapse
on a single curve in a wide $T$ range up to twice the $T\mathrm{_{c}}$
suggesting a $T$-indpendent background response present also in the
SC state. We subtract this normal-state background response to obtain
the SC-state response, as shown in Fig. \ref{fig:DrvsT} (c) and (d).
At the low fluence the subtraction does not significantly change the
shape of the response while at the high fluence it leads to the complete
removal of the sub-picosecond timescale dynamics, which is associated
with the normal state response, justifying the subtraction procedure. 

The SC-response shows a $\sim0.5$ ps risetime followed by $\sim5$
ps decay time at the low excitation. At the high excitation the risetime
is faster on $\sim0.2$ ps timescale while the relaxation slows down
to nanosecond timescale. In both cases the relaxation slows down when
approaching $T_{\mathrm{c}}$ from below.

\section{Discussion}

\begin{figure}
\includegraphics[angle=-90,width=1\columnwidth]{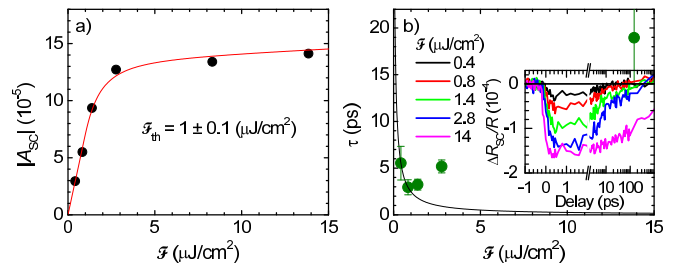}

\caption{(a) The fluence dependence of the transient amplitude at $T=7$ K.
The thin line in is the saturation model\cite{KusarKabanov2008} fit
discussed in text. (b) The dominant relaxation time in the superconducting
state ($T=7$ K). The thin line in is the Rothwarf-Tyalor fit\cite{KabanovDemsar2005}.
The inset to (b) shows the 7-K transients with the 32-K normal state
response subtracted.}
\label{fig:AvsF} 
\end{figure}

\subsection{Excitation density dependence}

The saturation behaviour of the transient-reflectivity amplitude with
increasing excitation density was observed in a number of gapped systems
such as superconductors\cite{KusarKabanov2008,GiannettiCoslovich2009,MerteljKusar2010,StojchevskaKusar2011,CoslovichGiannetti2011,BeyerStaedter2011,StojchevskaMertelj2012}
and charge-density wave compounds\cite{TomeljakSchaefer2009,YusupovMertelj2010}.
The saturation of the transient reflectivity amplitude was associated
with a nonthermal destruction of the condensate and complete closure
of the gap. Due to an inhomogeneous excitation resulting from a finite
light penetration depth and finite beam diameters the exact shape
of the amplitude versus $\mathcal{F}$ curve depends on geometrical
parameters. In order to obtain the bulk SC state destruction energy
density, $U_{\mathrm{d}}$, we therefore use the inhomogeneous SC-state
destruction model\cite{KusarKabanov2008} to fit the fluence dependence
of the transient reflectivity amplitude and determine the external
SC state destruction threshold fluence, $\mathcal{F}_{\mathrm{th}}=1\pm0.1$
$\mu$J/cm$^{2}$ {[}see Fig. \ref{fig:AvsF} (a){]}. The bulk SC
state destruction energy density required to completely destroy the
superconducting state is then obtained as $U_{\mathrm{d}}/k_{\mathrm{B}}=\mathcal{F}_{\mathrm{th}}(1-R)/\lambda_{\mathrm{p}}=0.68$
K/Fe, where $\lambda_{\mathrm{p}}=34$ nm is the light penetration
depth and $R=0.37$ the reflectivity at 1.55-eV photon energy taken
from data\cite{BarisicWu2010} on related Co doped Ba-122. As previously
noted\cite{StojchevskaMertelj2012} this value is much smaller than
the energy needed to heat the sample thermally above $T_{C}$ indicating
that the SC destruction is highly nonthermal. 

\begin{figure}
\includegraphics[angle=-90,width=0.8\columnwidth]{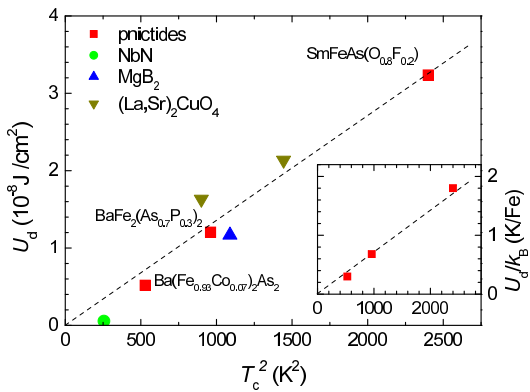}

\caption{The in-plane\cite{planar} SC-state optical destruction energy density
as function of $T\mathrm{_{c}^{2}}$ for some iron based pnictides
compared to NbN\cite{BeckKlammer2011}, MgB$_{2}$\cite{DemsarAveritt2003}
and (La,Sr)$_{2}$CuO$_{4}$\cite{KusarKabanov2008}. The inset shows
$U\mathrm{_{d}}$ in pnictides normalized to the Fe content.}
\label{fig:UpvTc}
\end{figure}

In Fig. \ref{fig:UpvTc} we compare the SC-condensate optical destruction
energy in iron pnictides with some other superconductors. As discussed
previously\cite{StojchevskaKusar2011} $U_{\mathrm{d}}$ is roughly
proportional to $T_{C}^{2}$. The actual value for each compound depends
on the thermodynamic condensation energy and the amount of the energy
lost by transfer to the non-pair-breaking subgap phonons\cite{StojchevskaKusar2011}.
It is therefore somewhat surprising that the in-plane\cite{planar}
destruction-energy densities for very different layered compounds
lie rather close to the same line with the exception of the three
dimensional NbN.

Within the iron-pnictide class the accuracy of the scaling with $T_{C}^{2}$
is enhanced if one considers the optical destruction energy normalized
to the Fe content (see inset to Fig. \ref{fig:UpvTc}). This suggests
that the differences of the detailed gap structure%
\footnote{For example the presence of nodes in BaFe$_{2}$(As,P)$_{2}$.%
} between different members contribute only a small correction to the
free-energy gain in the SC state.

\subsection{Temperature dependence and the SC gap}

\begin{figure}
\bigskip{}
\includegraphics[angle=-90,width=1\columnwidth]{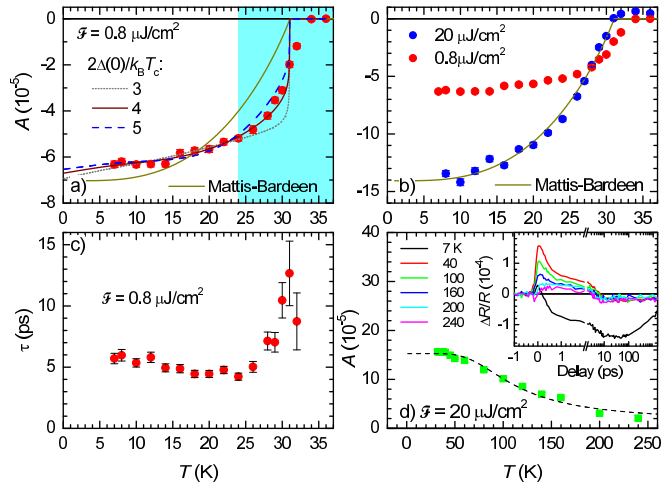}

\caption{The temperature dependence of the SC response amplitude for low (a)
and high (b) excitation density. The lines are fits discussed in text,
where the shaded region in (a) was excluded from the bottleneck model
fits. (c) The low-excitation SC-response relaxation time as a function
of temperature. (d) Temperature dependence of the normal-state transient
reflectivity amplitude. The dashed line is a bottleneck fit with $T$-independent
gap. Transients at a few characteristic temperatures are shown in
the inset. }
\label{fig:AvsT}
\end{figure}

Far above the saturation $\mathcal{F}$ the transient reflectivity
on short timescales can be understood as the difference between the
SC and normal state reflectivities and can be described in terms of
the high frequency limit of the Mattis-Bardeen formula\cite{MattisBardeen1958}: 

\begin{equation}
\Delta R_{\mathrm{SC}}\varpropto\left(\frac{\triangle(T)}{\hbar\omega}\right)^{2}\ln\left(\frac{3.3\hbar\omega}{\triangle(T)}\right)
\end{equation}
where $\hbar\omega$ is the photon energy and $\triangle(T)$ is a
temperature dependent gap. Using the BCS temperature dependent gap\cite{calib}
the formula fits well the temperature dependence of the transient
reflectivity amplitude measured in the strong excitation regime as
shown in Fig. \ref{fig:AvsT} (b). 

In the weak perturbation limit the transient optical response of superconductors
was shown to be governed by the phonon bottleneck effects.\cite{RothwarfTaylor1967,KabanovDemsar99,KabanovDemsar2005,TorchinskyMcIver2011}
The dynamics is usually discussed in the framework of simple effective
two-electronic-level models such is the Rothwarf-Taylor model\cite{RothwarfTaylor1967}
and the related Kabanov bottleneck model\cite{KabanovDemsar99}. Both
enable determination of the characteristic SC gap energy from the
$T$-dependent transient response amplitude.\cite{KabanovDemsar99,KabanovDemsar2005}

The weak excitation limit is difficult to achieve in the case of iron
pnictides due to rather small transient reflectivity magnitudes and
low optical destruction thresholds. In the present case the lowest
fluence of 0.8 $\mu$J/cm$^{2}$ used for $T$-scans appears to be
near the limit of the $\mathcal{F}$-linear response region. The $\mathcal{F}$-dependent
dominant%
\footnote{The relaxation at higher $\mathcal{F}$ is clearly non exponential
where the faster part can be associated with the Rothwarf-Taylor bottleneck,
while the slower one corresponds to the cooling after all degrees
of freedom have been thermalized.\cite{MerteljKusar2010} At excitation
densities above a few $\mu$J/cm$^{2}$ the amount of absorbed optical
energy is large enough to heat the phonon bath above $T\mathrm{_{c}}$
and the SC state recovery is governed by the energy escape from the
probed volume.%
} relaxation time shown in Fig. \ref{fig:AvsF} (b) shows the expected
Rothwarf-Taylor behaviour,\cite{KabanovDemsar2005} $\tau=\tau_{0}(\mathcal{F}+\mathcal{F}_{\mathrm{T}})^{-1}$,
below $\mathcal{F}\sim0.8$ $\mu$J/cm$^{2}$ only, with $\mathcal{F}_{T}=0.05$
$\mu$J/cm$^{2}$ at 7K, where $\mathcal{F}_{\mathrm{T}}$ corresponds
to the fluence at which the density of the photoexcited quasiparticles
is comparable to the thermally excited quasiparticle density.

On the other hand, the amplitude of the response appears almost linear
up to $\mathcal{F}\sim1.4$ $\mu$J/cm$^{2}$ at the lowest $T$ while
the $\mbox{\ensuremath{\mathcal{F}}=}0.8$-$\mu$J/cm$^{2}$ amplitude
reaches the saturation value only above $T\sim28$K {[}see Fig. \ref{fig:AvsT}
(b){]}. We therefore assume that for $T\lesssim25K$ the response
is $\mathcal{F}$-linear and the $T$-dependent amplitude, $A(T)$,
is proportional to the low-$\mathcal{F}$ limit and follows the bottleneck
model from Kabanov \textit{et al.}\cite{KabanovDemsar99},\begin{widetext}

\begin{equation}
A(T)\propto n_{\mathrm{pe}}\propto\left[\left(\frac{2\Delta\left(T\right)}{k_{\mathrm{B}}T_{\mathrm{c}}}+\frac{T}{T\mathrm{_{c}}}\right)\left(1+g_{\mathrm{ph}}\sqrt{\frac{k_{\mathrm{B}}T}{\Delta\left(T\right)}}\exp\left(-\frac{\Delta\left(T\right)}{k_{\mathrm{B}}T}\right)\right)\right]^{-1},\label{eq:npevsT}
\end{equation}
\end{widetext}where $g_{\mathrm{ph}}$ represents the relative effective
number of the involved phonon degrees of freedom. Fitting $A(T)$
in Fig. \ref{fig:AvsT} (a) using (\ref{eq:npevsT}) with the BCS
temperature dependent gap, $\Delta(T)$, we obtain $2\mbox{\ensuremath{\Delta}}(0)/k_{\textrm{B}}T\mathrm{_{c}}=4\pm1$,
with $\Delta(0)\sim5$ meV. This value is consistent with the gap
on the $\gamma$ hole Fermi surface and smaller than the gap on the
electron Fermi surfaces as obtained by ARPES\cite{zhangYe2012} in
the samples from the same batch. The relaxation dynamics detected
by means of 1.55-eV probe photons in BaFe$_{2}$(As$_{1-x}$P$_{x}$)$_{2}$
can therefore be attributed to the hole Fermi surfaces as suggested\cite{TorchinskyMcIver2011}
for K and Co doped Ba-122.

Despite the presence of several bands with different gaps in the case
of iron based pnictides and the presence of the gap node\cite{zhangYe2012}
on the $\alpha$ hole Fermi surface in BaFe$_{2}$(As$_{1-x}$P$_{x}$)$_{2}$
the amplitude of the weak-excitation response in the SC state seems
to be reasonably well described by the bottleneck model similarly
to the cuprate superconductors.\cite{KabanovDemsar2005}

The weak-excitation divergence of the relaxation time of the SC signal
at $T_{\mathrm{c}}$ {[}see Fig. \ref{fig:AvsT} (c){]} can also be
well described by the bottleneck model.\cite{KabanovDemsar99} In
the proximity to the transition temperature the SC gap becomes smaller,
which means that the probability to find a boson with the energy higher
that the gap size becomes higher and this slows down the relaxation
of photoexcited quasiparticles. The low-T divergence, predicted by
the Rothwarf-Taylor model,\cite{KabanovDemsar2005} is, on the other
hand, cut off by the rather high excitation density $\mathcal{F}\gg\mathcal{F}_{\mathrm{T}}$.

At low excitation fluence we observe a measurable response up to 2
K above $T_{\mathrm{c}}=30$ K, which vanishes at the highest fluence.
We attribute this response to SC fluctuations\cite{MadanKurosawa2014}
although we can not rule out a $+1$ K error in the determination
of the sample $T$\cite{calib}. This error does not significantly
affect the gap determination from the fit (\ref{eq:npevsT}) above.

The normal state sub-ps transient response, which shows a vanishing
amplitude with increasing temperature disappearing around $T=200$
K, is very similar to the behaviour in Co-doped Ba-122\cite{StojchevskaMertelj2012},
where it was associated with the presence of the pseudogap due to
the nematic fluctuations. It is plausible to assume that the origin
is the same in the present case. The temperature range of vanishing
pseudogap is about 50 K above the pseudogap formation temperature
based on ARPES results\cite{ShimojimaSonobe2014}. Contrary to Co-doped
Ba-122 it does not show any 4-fold axis symmetry breaking. Since a
global external symmetry breaking (strain) field is necessary to orient
the fluctuations across the whole experimental volume along one direction
this observation indicates the absence of a global field, but it does
not rule out local fields with varying orientation enhancing the nematic
fluctuations. 

Using a bottleneck fit with a $T$-indpendent gap as in Ref. {[}\onlinecite{StojchevskaMertelj2012}{]}
we obtain the characteristic pseudogap magnitude, $2\Delta_{\mathrm{PG}}=690\pm70$
K, which is within the errorbars identical to the magnitude ($2\Delta_{\mathrm{PG}}=660\pm100$
K) in the optimally Co doped Ba-122.\cite{StojchevskaMertelj2012}

\section{Conclusions}

Conducting a systematic wide-fluence-range optical time-resolved optical
pump-probe study in optimally isovalently doped BaFe$_{2}$(As$_{0.7}$P$_{0.3}$)$_{2}$
we find a similar behaviour to previously studied\cite{TorchinskyMcIver2011,StojchevskaMertelj2012}
Ba(Fe,Co)$_{2}$As$_{2}$.

The superconducting gap magnitude of $2\mbox{\ensuremath{\Delta}}(0)/k_{\textrm{B}}T\mathrm{_{c}}=4\pm1$
is consistent with the gap on the $\gamma$ hole Fermi surface. The
relaxation dynamics detected by means of 1.55-eV probe photons in
Ba-122 can therefore be attributed to the hole Fermi surfaces as suggested
previously by Torchinsky \emph{et al.} {[}\onlinecite{TorchinskyMcIver2011}{]}.

The normal state response indicates the presence of a pseudogap related
to the nematic fluctuations up to $T\sim200$ K as generally observed
in the electron doped iron based pnictide superconductors. 

The in-plane\cite{planar} nonthermal optical-destruction volume energy
densities are found to scale linearly with $T\mathrm{_{c}^{2}}$ and
lie close to the same line for a range of different layered superconductors. 
\begin{acknowledgments}
Work at Jozef Stefan Institute was supported by ARRS (Grant No. P1-0040). 
\end{acknowledgments}
\bibliography{biblio}

\begin{thebibliography}{27}%
\makeatletter
\providecommand \@ifxundefined [1]{%
 \@ifx{#1\undefined}
}%
\providecommand \@ifnum [1]{%
 \ifnum #1\expandafter \@firstoftwo
 \else \expandafter \@secondoftwo
 \fi
}%
\providecommand \@ifx [1]{%
 \ifx #1\expandafter \@firstoftwo
 \else \expandafter \@secondoftwo
 \fi
}%
\providecommand \natexlab [1]{#1}%
\providecommand \enquote  [1]{``#1''}%
\providecommand \bibnamefont  [1]{#1}%
\providecommand \bibfnamefont [1]{#1}%
\providecommand \citenamefont [1]{#1}%
\providecommand \href@noop [0]{\@secondoftwo}%
\providecommand \href [0]{\begingroup \@sanitize@url \@href}%
\providecommand \@href[1]{\@@startlink{#1}\@@href}%
\providecommand \@@href[1]{\endgroup#1\@@endlink}%
\providecommand \@sanitize@url [0]{\catcode `\\12\catcode `\$12\catcode
  `\&12\catcode `\#12\catcode `\^12\catcode `\_12\catcode `\%12\relax}%
\providecommand \@@startlink[1]{}%
\providecommand \@@endlink[0]{}%
\providecommand \url  [0]{\begingroup\@sanitize@url \@url }%
\providecommand \@url [1]{\endgroup\@href {#1}{\urlprefix }}%
\providecommand \urlprefix  [0]{URL }%
\providecommand \Eprint [0]{\href }%
\providecommand \doibase [0]{http://dx.doi.org/}%
\providecommand \selectlanguage [0]{\@gobble}%
\providecommand \bibinfo  [0]{\@secondoftwo}%
\providecommand \bibfield  [0]{\@secondoftwo}%
\providecommand \translation [1]{[#1]}%
\providecommand \BibitemOpen [0]{}%
\providecommand \bibitemStop [0]{}%
\providecommand \bibitemNoStop [0]{.\EOS\space}%
\providecommand \EOS [0]{\spacefactor3000\relax}%
\providecommand \BibitemShut  [1]{\csname bibitem#1\endcsname}%
\let\auto@bib@innerbib\@empty
%</preamble>
\bibitem [{\citenamefont {Zhang}\ \emph {et~al.}(2012)\citenamefont {Zhang},
  \citenamefont {Ye}, \citenamefont {Ge}, \citenamefont {Chen}, \citenamefont
  {Jiang}, \citenamefont {Xu}, \citenamefont {Xie},\ and\ \citenamefont
  {Feng}}]{zhangYe2012}%
  \BibitemOpen
  \bibfield  {author} {\bibinfo {author} {\bibfnamefont {Y.}~\bibnamefont
  {Zhang}}, \bibinfo {author} {\bibfnamefont {Z.}~\bibnamefont {Ye}}, \bibinfo
  {author} {\bibfnamefont {Q.}~\bibnamefont {Ge}}, \bibinfo {author}
  {\bibfnamefont {F.}~\bibnamefont {Chen}}, \bibinfo {author} {\bibfnamefont
  {J.}~\bibnamefont {Jiang}}, \bibinfo {author} {\bibfnamefont
  {M.}~\bibnamefont {Xu}}, \bibinfo {author} {\bibfnamefont {B.}~\bibnamefont
  {Xie}}, \ and\ \bibinfo {author} {\bibfnamefont {D.}~\bibnamefont {Feng}},\
  }\href@noop {} {\bibfield  {journal} {\bibinfo  {journal} {Nature Physics}\
  }\textbf {\bibinfo {volume} {8}},\ \bibinfo {pages} {371} (\bibinfo {year}
  {2012})}\BibitemShut {NoStop}%
\bibitem [{\citenamefont {Terashima}\ \emph {et~al.}(2009)\citenamefont
  {Terashima}, \citenamefont {Sekiba}, \citenamefont {Bowen}, \citenamefont
  {Nakayama}, \citenamefont {Kawahara}, \citenamefont {Sato}, \citenamefont
  {Richard}, \citenamefont {Xu}, \citenamefont {Li}, \citenamefont {Cao},
  \citenamefont {Xu}, \citenamefont {Ding},\ and\ \citenamefont
  {Takahashi}}]{TerashimaSekiba2009}%
  \BibitemOpen
  \bibfield  {author} {\bibinfo {author} {\bibfnamefont {K.}~\bibnamefont
  {Terashima}}, \bibinfo {author} {\bibfnamefont {Y.}~\bibnamefont {Sekiba}},
  \bibinfo {author} {\bibfnamefont {J.~H.}\ \bibnamefont {Bowen}}, \bibinfo
  {author} {\bibfnamefont {K.}~\bibnamefont {Nakayama}}, \bibinfo {author}
  {\bibfnamefont {T.}~\bibnamefont {Kawahara}}, \bibinfo {author}
  {\bibfnamefont {T.}~\bibnamefont {Sato}}, \bibinfo {author} {\bibfnamefont
  {P.}~\bibnamefont {Richard}}, \bibinfo {author} {\bibfnamefont {Y.-M.}\
  \bibnamefont {Xu}}, \bibinfo {author} {\bibfnamefont {L.~J.}\ \bibnamefont
  {Li}}, \bibinfo {author} {\bibfnamefont {G.~H.}\ \bibnamefont {Cao}},
  \bibinfo {author} {\bibfnamefont {Z.-A.}\ \bibnamefont {Xu}}, \bibinfo
  {author} {\bibfnamefont {H.}~\bibnamefont {Ding}}, \ and\ \bibinfo {author}
  {\bibfnamefont {T.}~\bibnamefont {Takahashi}},\ }\href {\doibase
  10.1073/pnas.0900469106} {\bibfield  {journal} {\bibinfo  {journal}
  {Proceedings of the National Academy of Sciences}\ }\textbf {\bibinfo
  {volume} {106}},\ \bibinfo {pages} {7330} (\bibinfo {year}
  {2009})}\BibitemShut {NoStop}%
\bibitem [{\citenamefont {Reid}\ \emph {et~al.}(2010)\citenamefont {Reid},
  \citenamefont {Tanatar}, \citenamefont {Luo}, \citenamefont {Shakeripour},
  \citenamefont {Doiron-Leyraud}, \citenamefont {Ni}, \citenamefont {Bud'ko},
  \citenamefont {Canfield}, \citenamefont {Prozorov},\ and\ \citenamefont
  {Taillefer}}]{ReidTanatar2010}%
  \BibitemOpen
  \bibfield  {author} {\bibinfo {author} {\bibfnamefont {J.-P.}\ \bibnamefont
  {Reid}}, \bibinfo {author} {\bibfnamefont {M.~A.}\ \bibnamefont {Tanatar}},
  \bibinfo {author} {\bibfnamefont {X.~G.}\ \bibnamefont {Luo}}, \bibinfo
  {author} {\bibfnamefont {H.}~\bibnamefont {Shakeripour}}, \bibinfo {author}
  {\bibfnamefont {N.}~\bibnamefont {Doiron-Leyraud}}, \bibinfo {author}
  {\bibfnamefont {N.}~\bibnamefont {Ni}}, \bibinfo {author} {\bibfnamefont
  {S.~L.}\ \bibnamefont {Bud'ko}}, \bibinfo {author} {\bibfnamefont {P.~C.}\
  \bibnamefont {Canfield}}, \bibinfo {author} {\bibfnamefont {R.}~\bibnamefont
  {Prozorov}}, \ and\ \bibinfo {author} {\bibfnamefont {L.}~\bibnamefont
  {Taillefer}},\ }\href {\doibase 10.1103/PhysRevB.82.064501} {\bibfield
  {journal} {\bibinfo  {journal} {Phys. Rev. B}\ }\textbf {\bibinfo {volume}
  {82}},\ \bibinfo {pages} {064501} (\bibinfo {year} {2010})}\BibitemShut
  {NoStop}%
\bibitem [{\citenamefont {Kim}\ \emph {et~al.}(2010)\citenamefont {Kim},
  \citenamefont {Hirschfeld}, \citenamefont {Stewart}, \citenamefont
  {Kasahara}, \citenamefont {Shibauchi}, \citenamefont {Terashima},\ and\
  \citenamefont {Matsuda}}]{KimHirschfeld2010}%
  \BibitemOpen
  \bibfield  {author} {\bibinfo {author} {\bibfnamefont {J.~S.}\ \bibnamefont
  {Kim}}, \bibinfo {author} {\bibfnamefont {P.~J.}\ \bibnamefont {Hirschfeld}},
  \bibinfo {author} {\bibfnamefont {G.~R.}\ \bibnamefont {Stewart}}, \bibinfo
  {author} {\bibfnamefont {S.}~\bibnamefont {Kasahara}}, \bibinfo {author}
  {\bibfnamefont {T.}~\bibnamefont {Shibauchi}}, \bibinfo {author}
  {\bibfnamefont {T.}~\bibnamefont {Terashima}}, \ and\ \bibinfo {author}
  {\bibfnamefont {Y.}~\bibnamefont {Matsuda}},\ }\href {\doibase
  10.1103/PhysRevB.81.214507} {\bibfield  {journal} {\bibinfo  {journal} {Phys.
  Rev. B}\ }\textbf {\bibinfo {volume} {81}},\ \bibinfo {pages} {214507}
  (\bibinfo {year} {2010})}\BibitemShut {NoStop}%
\bibitem [{\citenamefont {Rothwarf}\ and\ \citenamefont
  {Taylor}(1967)}]{RothwarfTaylor1967}%
  \BibitemOpen
  \bibfield  {author} {\bibinfo {author} {\bibfnamefont {A.}~\bibnamefont
  {Rothwarf}}\ and\ \bibinfo {author} {\bibfnamefont {B.~N.}\ \bibnamefont
  {Taylor}},\ }\href {\doibase 10.1103/PhysRevLett.19.27} {\bibfield  {journal}
  {\bibinfo  {journal} {Phys. Rev. Lett.}\ }\textbf {\bibinfo {volume} {19}},\
  \bibinfo {pages} {27} (\bibinfo {year} {1967})}\BibitemShut {NoStop}%
\bibitem [{\citenamefont {Kabanov}\ \emph {et~al.}(2005)\citenamefont
  {Kabanov}, \citenamefont {Demsar},\ and\ \citenamefont
  {Mihailovic}}]{KabanovDemsar2005}%
  \BibitemOpen
  \bibfield  {author} {\bibinfo {author} {\bibfnamefont {V.~V.}\ \bibnamefont
  {Kabanov}}, \bibinfo {author} {\bibfnamefont {J.}~\bibnamefont {Demsar}}, \
  and\ \bibinfo {author} {\bibfnamefont {D.}~\bibnamefont {Mihailovic}},\
  }\href {\doibase 10.1103/PhysRevLett.95.147002} {\bibfield  {journal}
  {\bibinfo  {journal} {Phys. Rev. Lett.}\ }\textbf {\bibinfo {volume} {95}},\
  \bibinfo {pages} {147002} (\bibinfo {year} {2005})}\BibitemShut {NoStop}%
\bibitem [{\citenamefont {Torchinsky}\ \emph {et~al.}(2011)\citenamefont
  {Torchinsky}, \citenamefont {McIver}, \citenamefont {Hsieh}, \citenamefont
  {Chen}, \citenamefont {Luo}, \citenamefont {Wang},\ and\ \citenamefont
  {Gedik}}]{TorchinskyMcIver2011}%
  \BibitemOpen
  \bibfield  {author} {\bibinfo {author} {\bibfnamefont {D.~H.}\ \bibnamefont
  {Torchinsky}}, \bibinfo {author} {\bibfnamefont {J.~W.}\ \bibnamefont
  {McIver}}, \bibinfo {author} {\bibfnamefont {D.}~\bibnamefont {Hsieh}},
  \bibinfo {author} {\bibfnamefont {G.~F.}\ \bibnamefont {Chen}}, \bibinfo
  {author} {\bibfnamefont {J.~L.}\ \bibnamefont {Luo}}, \bibinfo {author}
  {\bibfnamefont {N.~L.}\ \bibnamefont {Wang}}, \ and\ \bibinfo {author}
  {\bibfnamefont {N.}~\bibnamefont {Gedik}},\ }\href {\doibase
  10.1103/PhysRevB.84.104518} {\bibfield  {journal} {\bibinfo  {journal} {Phys.
  Rev. B}\ }\textbf {\bibinfo {volume} {84}},\ \bibinfo {pages} {104518}
  (\bibinfo {year} {2011})}\BibitemShut {NoStop}%
\bibitem [{\citenamefont {Stojchevska}\ \emph {et~al.}(2012)\citenamefont
  {Stojchevska}, \citenamefont {Mertelj}, \citenamefont {Chu}, \citenamefont
  {Fisher},\ and\ \citenamefont {Mihailovic}}]{StojchevskaMertelj2012}%
  \BibitemOpen
  \bibfield  {author} {\bibinfo {author} {\bibfnamefont {L.}~\bibnamefont
  {Stojchevska}}, \bibinfo {author} {\bibfnamefont {T.}~\bibnamefont
  {Mertelj}}, \bibinfo {author} {\bibfnamefont {J.}~\bibnamefont {Chu}},
  \bibinfo {author} {\bibfnamefont {I.}~\bibnamefont {Fisher}}, \ and\ \bibinfo
  {author} {\bibfnamefont {D.}~\bibnamefont {Mihailovic}},\ }\href@noop {}
  {\bibfield  {journal} {\bibinfo  {journal} {Physical Review B}\ }\textbf
  {\bibinfo {volume} {86}},\ \bibinfo {pages} {024519} (\bibinfo {year}
  {2012})}\BibitemShut {NoStop}%
\bibitem [{\citenamefont {Kusar}\ \emph {et~al.}(2008)\citenamefont {Kusar},
  \citenamefont {Kabanov}, \citenamefont {Demsar}, \citenamefont {Mertelj},
  \citenamefont {Sugai},\ and\ \citenamefont {Mihailovic}}]{KusarKabanov2008}%
  \BibitemOpen
  \bibfield  {author} {\bibinfo {author} {\bibfnamefont {P.}~\bibnamefont
  {Kusar}}, \bibinfo {author} {\bibfnamefont {V.}~\bibnamefont {Kabanov}},
  \bibinfo {author} {\bibfnamefont {J.}~\bibnamefont {Demsar}}, \bibinfo
  {author} {\bibfnamefont {T.}~\bibnamefont {Mertelj}}, \bibinfo {author}
  {\bibfnamefont {S.}~\bibnamefont {Sugai}}, \ and\ \bibinfo {author}
  {\bibfnamefont {D.}~\bibnamefont {Mihailovic}},\ }\href {\doibase
  10.1103/PhysRevLett.101.227001} {\bibfield  {journal} {\bibinfo  {journal}
  {Physical Review Letters}\ }\textbf {\bibinfo {volume} {101}},\ \bibinfo
  {pages} {227001} (\bibinfo {year} {2008})}\BibitemShut {NoStop}%
\bibitem [{\citenamefont {Giannetti}\ \emph {et~al.}(2009)\citenamefont
  {Giannetti}, \citenamefont {Coslovich}, \citenamefont {Cilento},
  \citenamefont {Ferrini}, \citenamefont {Eisaki}, \citenamefont {Kaneko},
  \citenamefont {Greven},\ and\ \citenamefont
  {Parmigiani}}]{GiannettiCoslovich2009}%
  \BibitemOpen
  \bibfield  {author} {\bibinfo {author} {\bibfnamefont {C.}~\bibnamefont
  {Giannetti}}, \bibinfo {author} {\bibfnamefont {G.}~\bibnamefont
  {Coslovich}}, \bibinfo {author} {\bibfnamefont {F.}~\bibnamefont {Cilento}},
  \bibinfo {author} {\bibfnamefont {G.}~\bibnamefont {Ferrini}}, \bibinfo
  {author} {\bibfnamefont {H.}~\bibnamefont {Eisaki}}, \bibinfo {author}
  {\bibfnamefont {N.}~\bibnamefont {Kaneko}}, \bibinfo {author} {\bibfnamefont
  {M.}~\bibnamefont {Greven}}, \ and\ \bibinfo {author} {\bibfnamefont
  {F.}~\bibnamefont {Parmigiani}},\ }\href {\doibase
  10.1103/PhysRevB.79.224502} {\bibfield  {journal} {\bibinfo  {journal} {Phys.
  Rev. B}\ }\textbf {\bibinfo {volume} {79}},\ \bibinfo {pages} {224502}
  (\bibinfo {year} {2009})}\BibitemShut {NoStop}%
\bibitem [{\citenamefont {Mertelj}\ \emph {et~al.}(2010)\citenamefont
  {Mertelj}, \citenamefont {Kusar}, \citenamefont {Kabanov}, \citenamefont
  {Stojchevska}, \citenamefont {Zhigadlo}, \citenamefont {Katrych},
  \citenamefont {Bukowski}, \citenamefont {Karpinski}, \citenamefont
  {Weyeneth},\ and\ \citenamefont {Mihailovic}}]{MerteljKusar2010}%
  \BibitemOpen
  \bibfield  {author} {\bibinfo {author} {\bibfnamefont {T.}~\bibnamefont
  {Mertelj}}, \bibinfo {author} {\bibfnamefont {P.}~\bibnamefont {Kusar}},
  \bibinfo {author} {\bibfnamefont {V.~V.}\ \bibnamefont {Kabanov}}, \bibinfo
  {author} {\bibfnamefont {L.}~\bibnamefont {Stojchevska}}, \bibinfo {author}
  {\bibfnamefont {N.~D.}\ \bibnamefont {Zhigadlo}}, \bibinfo {author}
  {\bibfnamefont {S.}~\bibnamefont {Katrych}}, \bibinfo {author} {\bibfnamefont
  {Z.}~\bibnamefont {Bukowski}}, \bibinfo {author} {\bibfnamefont
  {J.}~\bibnamefont {Karpinski}}, \bibinfo {author} {\bibfnamefont
  {S.}~\bibnamefont {Weyeneth}}, \ and\ \bibinfo {author} {\bibfnamefont
  {D.}~\bibnamefont {Mihailovic}},\ }\href {\doibase
  10.1103/PhysRevB.81.224504} {\bibfield  {journal} {\bibinfo  {journal} {Phys.
  Rev. B}\ }\textbf {\bibinfo {volume} {81}},\ \bibinfo {pages} {224504}
  (\bibinfo {year} {2010})}\BibitemShut {NoStop}%
\bibitem [{\citenamefont {Stojchevska}\ \emph {et~al.}(2011)\citenamefont
  {Stojchevska}, \citenamefont {Kusar}, \citenamefont {Mertelj}, \citenamefont
  {Kabanov}, \citenamefont {Toda}, \citenamefont {Yao},\ and\ \citenamefont
  {Mihailovic}}]{StojchevskaKusar2011}%
  \BibitemOpen
  \bibfield  {author} {\bibinfo {author} {\bibfnamefont {L.}~\bibnamefont
  {Stojchevska}}, \bibinfo {author} {\bibfnamefont {P.}~\bibnamefont {Kusar}},
  \bibinfo {author} {\bibfnamefont {T.}~\bibnamefont {Mertelj}}, \bibinfo
  {author} {\bibfnamefont {V.~V.}\ \bibnamefont {Kabanov}}, \bibinfo {author}
  {\bibfnamefont {Y.}~\bibnamefont {Toda}}, \bibinfo {author} {\bibfnamefont
  {X.}~\bibnamefont {Yao}}, \ and\ \bibinfo {author} {\bibfnamefont
  {D.}~\bibnamefont {Mihailovic}},\ }\href {\doibase
  10.1103/PhysRevB.84.180507} {\bibfield  {journal} {\bibinfo  {journal} {Phys.
  Rev. B}\ }\textbf {\bibinfo {volume} {84}},\ \bibinfo {pages} {180507}
  (\bibinfo {year} {2011})}\BibitemShut {NoStop}%
\bibitem [{\citenamefont {Coslovich}\ \emph {et~al.}(2011)\citenamefont
  {Coslovich}, \citenamefont {Giannetti}, \citenamefont {Cilento},
  \citenamefont {Dal~Conte}, \citenamefont {Ferrini}, \citenamefont
  {Galinetto}, \citenamefont {Greven}, \citenamefont {Eisaki}, \citenamefont
  {Raichle}, \citenamefont {Liang}, \citenamefont {Damascelli},\ and\
  \citenamefont {Parmigiani}}]{CoslovichGiannetti2011}%
  \BibitemOpen
  \bibfield  {author} {\bibinfo {author} {\bibfnamefont {G.}~\bibnamefont
  {Coslovich}}, \bibinfo {author} {\bibfnamefont {C.}~\bibnamefont
  {Giannetti}}, \bibinfo {author} {\bibfnamefont {F.}~\bibnamefont {Cilento}},
  \bibinfo {author} {\bibfnamefont {S.}~\bibnamefont {Dal~Conte}}, \bibinfo
  {author} {\bibfnamefont {G.}~\bibnamefont {Ferrini}}, \bibinfo {author}
  {\bibfnamefont {P.}~\bibnamefont {Galinetto}}, \bibinfo {author}
  {\bibfnamefont {M.}~\bibnamefont {Greven}}, \bibinfo {author} {\bibfnamefont
  {H.}~\bibnamefont {Eisaki}}, \bibinfo {author} {\bibfnamefont
  {M.}~\bibnamefont {Raichle}}, \bibinfo {author} {\bibfnamefont
  {R.}~\bibnamefont {Liang}}, \bibinfo {author} {\bibfnamefont
  {A.}~\bibnamefont {Damascelli}}, \ and\ \bibinfo {author} {\bibfnamefont
  {F.}~\bibnamefont {Parmigiani}},\ }\href {\doibase
  10.1103/PhysRevB.83.064519} {\bibfield  {journal} {\bibinfo  {journal} {Phys.
  Rev. B}\ }\textbf {\bibinfo {volume} {83}},\ \bibinfo {pages} {064519}
  (\bibinfo {year} {2011})}\BibitemShut {NoStop}%
\bibitem [{\citenamefont {Beyer}\ \emph {et~al.}(2011)\citenamefont {Beyer},
  \citenamefont {St\"adter}, \citenamefont {Beck}, \citenamefont {Sch\"afer},
  \citenamefont {Kabanov}, \citenamefont {Logvenov}, \citenamefont {Bozovic},
  \citenamefont {Koren},\ and\ \citenamefont {Demsar}}]{BeyerStaedter2011}%
  \BibitemOpen
  \bibfield  {author} {\bibinfo {author} {\bibfnamefont {M.}~\bibnamefont
  {Beyer}}, \bibinfo {author} {\bibfnamefont {D.}~\bibnamefont {St\"adter}},
  \bibinfo {author} {\bibfnamefont {M.}~\bibnamefont {Beck}}, \bibinfo {author}
  {\bibfnamefont {H.}~\bibnamefont {Sch\"afer}}, \bibinfo {author}
  {\bibfnamefont {V.~V.}\ \bibnamefont {Kabanov}}, \bibinfo {author}
  {\bibfnamefont {G.}~\bibnamefont {Logvenov}}, \bibinfo {author}
  {\bibfnamefont {I.}~\bibnamefont {Bozovic}}, \bibinfo {author} {\bibfnamefont
  {G.}~\bibnamefont {Koren}}, \ and\ \bibinfo {author} {\bibfnamefont
  {J.}~\bibnamefont {Demsar}},\ }\href {\doibase 10.1103/PhysRevB.83.214515}
  {\bibfield  {journal} {\bibinfo  {journal} {Phys. Rev. B}\ }\textbf {\bibinfo
  {volume} {83}},\ \bibinfo {pages} {214515} (\bibinfo {year}
  {2011})}\BibitemShut {NoStop}%
\bibitem [{\citenamefont {Tomeljak}\ \emph {et~al.}(2009)\citenamefont
  {Tomeljak}, \citenamefont {Sch\"afer}, \citenamefont {St\"adter},
  \citenamefont {Beyer}, \citenamefont {Biljakovic},\ and\ \citenamefont
  {Demsar}}]{TomeljakSchaefer2009}%
  \BibitemOpen
  \bibfield  {author} {\bibinfo {author} {\bibfnamefont {A.}~\bibnamefont
  {Tomeljak}}, \bibinfo {author} {\bibfnamefont {H.}~\bibnamefont {Sch\"afer}},
  \bibinfo {author} {\bibfnamefont {D.}~\bibnamefont {St\"adter}}, \bibinfo
  {author} {\bibfnamefont {M.}~\bibnamefont {Beyer}}, \bibinfo {author}
  {\bibfnamefont {K.}~\bibnamefont {Biljakovic}}, \ and\ \bibinfo {author}
  {\bibfnamefont {J.}~\bibnamefont {Demsar}},\ }\href {\doibase
  10.1103/PhysRevLett.102.066404} {\bibfield  {journal} {\bibinfo  {journal}
  {Phys. Rev. Lett.}\ }\textbf {\bibinfo {volume} {102}},\ \bibinfo {pages}
  {066404} (\bibinfo {year} {2009})}\BibitemShut {NoStop}%
\bibitem [{\citenamefont {Yusupov}\ \emph {et~al.}(2010)\citenamefont
  {Yusupov}, \citenamefont {Mertelj}, \citenamefont {Kusar}, \citenamefont
  {Kabanov}, \citenamefont {Brazovskii}, \citenamefont {Chu}, \citenamefont
  {Fisher},\ and\ \citenamefont {Mihailovic}}]{YusupovMertelj2010}%
  \BibitemOpen
  \bibfield  {author} {\bibinfo {author} {\bibfnamefont {R.~V.}\ \bibnamefont
  {Yusupov}}, \bibinfo {author} {\bibfnamefont {T.}~\bibnamefont {Mertelj}},
  \bibinfo {author} {\bibfnamefont {P.}~\bibnamefont {Kusar}}, \bibinfo
  {author} {\bibfnamefont {V.}~\bibnamefont {Kabanov}}, \bibinfo {author}
  {\bibfnamefont {S.}~\bibnamefont {Brazovskii}}, \bibinfo {author}
  {\bibfnamefont {J.-H.}\ \bibnamefont {Chu}}, \bibinfo {author} {\bibfnamefont
  {I.~R.}\ \bibnamefont {Fisher}}, \ and\ \bibinfo {author} {\bibfnamefont
  {D.}~\bibnamefont {Mihailovic}},\ }\href {\doibase
  http://dx.doi.org/10.1038/nphys1738} {\bibfield  {journal} {\bibinfo
  {journal} {Nature Physics}\ }\textbf {\bibinfo {volume} {6}},\ \bibinfo
  {pages} {681} (\bibinfo {year} {2010})}\BibitemShut {NoStop}%
\bibitem [{\citenamefont {Bari\ifmmode \check{s}\else
  \v{s}\fi{}i\ifmmode~\acute{c}\else \'{c}\fi{}}\ \emph
  {et~al.}(2010)\citenamefont {Bari\ifmmode \check{s}\else
  \v{s}\fi{}i\ifmmode~\acute{c}\else \'{c}\fi{}}, \citenamefont {Wu},
  \citenamefont {Dressel}, \citenamefont {Li}, \citenamefont {Cao},\ and\
  \citenamefont {Xu}}]{BarisicWu2010}%
  \BibitemOpen
  \bibfield  {author} {\bibinfo {author} {\bibfnamefont {N.}~\bibnamefont
  {Bari\ifmmode \check{s}\else \v{s}\fi{}i\ifmmode~\acute{c}\else \'{c}\fi{}}},
  \bibinfo {author} {\bibfnamefont {D.}~\bibnamefont {Wu}}, \bibinfo {author}
  {\bibfnamefont {M.}~\bibnamefont {Dressel}}, \bibinfo {author} {\bibfnamefont
  {L.~J.}\ \bibnamefont {Li}}, \bibinfo {author} {\bibfnamefont {G.~H.}\
  \bibnamefont {Cao}}, \ and\ \bibinfo {author} {\bibfnamefont {Z.~A.}\
  \bibnamefont {Xu}},\ }\href {\doibase 10.1103/PhysRevB.82.054518} {\bibfield
  {journal} {\bibinfo  {journal} {Phys. Rev. B}\ }\textbf {\bibinfo {volume}
  {82}},\ \bibinfo {pages} {054518} (\bibinfo {year} {2010})}\BibitemShut
  {NoStop}%
\bibitem [{pla()}]{planar}%
  \BibitemOpen
  \href@noop {} {}\bibinfo {note} {We normalize the energy density to a single
  FeAs , CuO$_{2}$, Mg or Nd plane by multiplying the volume density by the
  interplane distances.}\BibitemShut {Stop}%
\bibitem [{Note1()}]{Note1}%
  \BibitemOpen
  \bibinfo {note} {For example the presence of nodes in
  BaFe$_{2}$(As,P)$_{2}$.}\BibitemShut {Stop}%
\bibitem [{\citenamefont {Beck}\ \emph {et~al.}(2011)\citenamefont {Beck},
  \citenamefont {Klammer}, \citenamefont {Lang}, \citenamefont {Leiderer},
  \citenamefont {Kabanov}, \citenamefont {Gol'tsman},\ and\ \citenamefont
  {Demsar}}]{BeckKlammer2011}%
  \BibitemOpen
  \bibfield  {author} {\bibinfo {author} {\bibfnamefont {M.}~\bibnamefont
  {Beck}}, \bibinfo {author} {\bibfnamefont {M.}~\bibnamefont {Klammer}},
  \bibinfo {author} {\bibfnamefont {S.}~\bibnamefont {Lang}}, \bibinfo {author}
  {\bibfnamefont {P.}~\bibnamefont {Leiderer}}, \bibinfo {author}
  {\bibfnamefont {V.~V.}\ \bibnamefont {Kabanov}}, \bibinfo {author}
  {\bibfnamefont {G.~N.}\ \bibnamefont {Gol'tsman}}, \ and\ \bibinfo {author}
  {\bibfnamefont {J.}~\bibnamefont {Demsar}},\ }\href {\doibase
  10.1103/PhysRevLett.107.177007} {\bibfield  {journal} {\bibinfo  {journal}
  {Phys. Rev. Lett.}\ }\textbf {\bibinfo {volume} {107}},\ \bibinfo {pages}
  {177007} (\bibinfo {year} {2011})}\BibitemShut {NoStop}%
\bibitem [{\citenamefont {Demsar}\ \emph {et~al.}(2003)\citenamefont {Demsar},
  \citenamefont {Averitt}, \citenamefont {Taylor}, \citenamefont {Kang},
  \citenamefont {Kim}, \citenamefont {Choi},\ and\ \citenamefont
  {Lee}}]{DemsarAveritt2003}%
  \BibitemOpen
  \bibfield  {author} {\bibinfo {author} {\bibfnamefont {J.}~\bibnamefont
  {Demsar}}, \bibinfo {author} {\bibfnamefont {R.~D.}\ \bibnamefont {Averitt}},
  \bibinfo {author} {\bibfnamefont {A.~J.}\ \bibnamefont {Taylor}}, \bibinfo
  {author} {\bibfnamefont {W.-N.}\ \bibnamefont {Kang}}, \bibinfo {author}
  {\bibfnamefont {H.~J.}\ \bibnamefont {Kim}}, \bibinfo {author} {\bibfnamefont
  {E.-M.}\ \bibnamefont {Choi}}, \ and\ \bibinfo {author} {\bibfnamefont
  {S.-I.}\ \bibnamefont {Lee}},\ }\href {\doibase 10.1142/S0217979203021605}
  {\bibfield  {journal} {\bibinfo  {journal} {International Journal of Modern
  Physics B}\ }\textbf {\bibinfo {volume} {17}},\ \bibinfo {pages} {3675}
  (\bibinfo {year} {2003})}\BibitemShut {NoStop}%
\bibitem [{\citenamefont {Mattis}\ and\ \citenamefont
  {Bardeen}(1958)}]{MattisBardeen1958}%
  \BibitemOpen
  \bibfield  {author} {\bibinfo {author} {\bibfnamefont {D.~C.}\ \bibnamefont
  {Mattis}}\ and\ \bibinfo {author} {\bibfnamefont {J.}~\bibnamefont
  {Bardeen}},\ }\href {\doibase 10.1103/PhysRev.111.412} {\bibfield  {journal}
  {\bibinfo  {journal} {Phys. Rev.}\ }\textbf {\bibinfo {volume} {111}},\
  \bibinfo {pages} {412} (\bibinfo {year} {1958})}\BibitemShut {NoStop}%
\bibitem [{cal()}]{calib}%
  \BibitemOpen
  \href@noop {} {}\bibinfo {note} {From the fit $T_{\mathrm{c}}=31\pm0.2$ K was
  obtained which is 1 K higher than the maximum bulk $T_{\mathrm{c}} $ in this
  system. The difference can not be attributed to the thermal gradient due to
  the laser excitation and the room radiation thermal load since it should
  decrease the apparent $T_{\mathrm{c}} $, but to an error in the cryostat $T$
  calibration, which was checked a posteriori by means of a calibrated diode
  temperature sensor mounted at the sample position to be $\pm1$
  K.}\BibitemShut {Stop}%
\bibitem [{\citenamefont {Kabanov}\ \emph {et~al.}(1999)\citenamefont
  {Kabanov}, \citenamefont {Demsar}, \citenamefont {Podobnik},\ and\
  \citenamefont {Mihailovic}}]{KabanovDemsar99}%
  \BibitemOpen
  \bibfield  {author} {\bibinfo {author} {\bibfnamefont {V.~V.}\ \bibnamefont
  {Kabanov}}, \bibinfo {author} {\bibfnamefont {J.}~\bibnamefont {Demsar}},
  \bibinfo {author} {\bibfnamefont {B.}~\bibnamefont {Podobnik}}, \ and\
  \bibinfo {author} {\bibfnamefont {D.}~\bibnamefont {Mihailovic}},\ }\href
  {\doibase 10.1103/PhysRevB.59.1497} {\bibfield  {journal} {\bibinfo
  {journal} {Phys. Rev. B}\ }\textbf {\bibinfo {volume} {59}},\ \bibinfo
  {pages} {1497} (\bibinfo {year} {1999})}\BibitemShut {NoStop}%
\bibitem [{Note2()}]{Note2}%
  \BibitemOpen
  \bibinfo {note} {The relaxation at higher $\protect \mathcal {F}$ is clearly
  non exponential where the faster part can be associated with the
  Rothwarf-Taylor bottleneck, while the slower one corresponds to the cooling
  after all degrees of freedom have been thermalized.\cite {MerteljKusar2010}
  At excitation densities above a few $\mu $J/cm$^{2}$ the amount of absorbed
  optical energy is large enough to heat the phonon bath above $T\protect
  \mathrm {_{c}}$ and the SC state recovery is governed by the energy escape
  from the probed volume.}\BibitemShut {Stop}%
\bibitem [{\citenamefont {Madan}\ \emph {et~al.}(2014)\citenamefont {Madan},
  \citenamefont {Kurosawa}, \citenamefont {Toda}, \citenamefont {Oda},
  \citenamefont {Mertelj}, \citenamefont {Kusar},\ and\ \citenamefont
  {Mihailovic}}]{MadanKurosawa2014}%
  \BibitemOpen
  \bibfield  {author} {\bibinfo {author} {\bibfnamefont {I.}~\bibnamefont
  {Madan}}, \bibinfo {author} {\bibfnamefont {T.}~\bibnamefont {Kurosawa}},
  \bibinfo {author} {\bibfnamefont {Y.}~\bibnamefont {Toda}}, \bibinfo {author}
  {\bibfnamefont {M.}~\bibnamefont {Oda}}, \bibinfo {author} {\bibfnamefont
  {T.}~\bibnamefont {Mertelj}}, \bibinfo {author} {\bibfnamefont
  {P.}~\bibnamefont {Kusar}}, \ and\ \bibinfo {author} {\bibfnamefont
  {D.}~\bibnamefont {Mihailovic}},\ }\href
  {http://dx.doi.org/10.1038/srep05656} {\bibfield  {journal} {\bibinfo
  {journal} {Sci. Rep.}\ }\textbf {\bibinfo {volume} {4}},\  (\bibinfo {year}
  {2014})}\BibitemShut {NoStop}%
\bibitem [{\citenamefont {Shimojima}\ \emph {et~al.}(2014)\citenamefont
  {Shimojima}, \citenamefont {Sonobe}, \citenamefont {Malaeb}, \citenamefont
  {Shinada}, \citenamefont {Chainani}, \citenamefont {Shin}, \citenamefont
  {Yoshida}, \citenamefont {Ideta}, \citenamefont {Fujimori}, \citenamefont
  {Kumigashira}, \citenamefont {Ono}, \citenamefont {Nakashima}, \citenamefont
  {Anzai}, \citenamefont {Arita}, \citenamefont {Ino}, \citenamefont
  {Namatame}, \citenamefont {Taniguchi}, \citenamefont {Nakajima},
  \citenamefont {Uchida}, \citenamefont {Tomioka}, \citenamefont {Ito},
  \citenamefont {Kihou}, \citenamefont {Lee}, \citenamefont {Iyo},
  \citenamefont {Eisaki}, \citenamefont {Ohgushi}, \citenamefont {Kasahara},
  \citenamefont {Terashima}, \citenamefont {Ikeda}, \citenamefont {Shibauchi},
  \citenamefont {Matsuda},\ and\ \citenamefont
  {Ishizaka}}]{ShimojimaSonobe2014}%
  \BibitemOpen
  \bibfield  {author} {\bibinfo {author} {\bibfnamefont {T.}~\bibnamefont
  {Shimojima}}, \bibinfo {author} {\bibfnamefont {T.}~\bibnamefont {Sonobe}},
  \bibinfo {author} {\bibfnamefont {W.}~\bibnamefont {Malaeb}}, \bibinfo
  {author} {\bibfnamefont {K.}~\bibnamefont {Shinada}}, \bibinfo {author}
  {\bibfnamefont {A.}~\bibnamefont {Chainani}}, \bibinfo {author}
  {\bibfnamefont {S.}~\bibnamefont {Shin}}, \bibinfo {author} {\bibfnamefont
  {T.}~\bibnamefont {Yoshida}}, \bibinfo {author} {\bibfnamefont
  {S.}~\bibnamefont {Ideta}}, \bibinfo {author} {\bibfnamefont
  {A.}~\bibnamefont {Fujimori}}, \bibinfo {author} {\bibfnamefont
  {H.}~\bibnamefont {Kumigashira}}, \bibinfo {author} {\bibfnamefont
  {K.}~\bibnamefont {Ono}}, \bibinfo {author} {\bibfnamefont {Y.}~\bibnamefont
  {Nakashima}}, \bibinfo {author} {\bibfnamefont {H.}~\bibnamefont {Anzai}},
  \bibinfo {author} {\bibfnamefont {M.}~\bibnamefont {Arita}}, \bibinfo
  {author} {\bibfnamefont {A.}~\bibnamefont {Ino}}, \bibinfo {author}
  {\bibfnamefont {H.}~\bibnamefont {Namatame}}, \bibinfo {author}
  {\bibfnamefont {M.}~\bibnamefont {Taniguchi}}, \bibinfo {author}
  {\bibfnamefont {M.}~\bibnamefont {Nakajima}}, \bibinfo {author}
  {\bibfnamefont {S.}~\bibnamefont {Uchida}}, \bibinfo {author} {\bibfnamefont
  {Y.}~\bibnamefont {Tomioka}}, \bibinfo {author} {\bibfnamefont
  {T.}~\bibnamefont {Ito}}, \bibinfo {author} {\bibfnamefont {K.}~\bibnamefont
  {Kihou}}, \bibinfo {author} {\bibfnamefont {C.~H.}\ \bibnamefont {Lee}},
  \bibinfo {author} {\bibfnamefont {A.}~\bibnamefont {Iyo}}, \bibinfo {author}
  {\bibfnamefont {H.}~\bibnamefont {Eisaki}}, \bibinfo {author} {\bibfnamefont
  {K.}~\bibnamefont {Ohgushi}}, \bibinfo {author} {\bibfnamefont
  {S.}~\bibnamefont {Kasahara}}, \bibinfo {author} {\bibfnamefont
  {T.}~\bibnamefont {Terashima}}, \bibinfo {author} {\bibfnamefont
  {H.}~\bibnamefont {Ikeda}}, \bibinfo {author} {\bibfnamefont
  {T.}~\bibnamefont {Shibauchi}}, \bibinfo {author} {\bibfnamefont
  {Y.}~\bibnamefont {Matsuda}}, \ and\ \bibinfo {author} {\bibfnamefont
  {K.}~\bibnamefont {Ishizaka}},\ }\href {\doibase 10.1103/PhysRevB.89.045101}
  {\bibfield  {journal} {\bibinfo  {journal} {Phys. Rev. B}\ }\textbf {\bibinfo
  {volume} {89}},\ \bibinfo {pages} {045101} (\bibinfo {year}
  {2014})}\BibitemShut {NoStop}%
\end{thebibliography}%

\end{document}